# A Functioning Beta Solution to the Challenge of Opening Transit Payment System Transaction Data

David Ory
Metropolitan Transportation Commission
San Francisco, CA, USA
DOry@mtc.ca.gov

Stephen Granger-Bevan
Formerly of Metropolitan Transportation Commission
San Francisco, CA, USA
stephen.grangerbevan.sjsu@gmail.com

## ABSTRACT

The deployment of smart-card-based public transit fare payment systems provides government the opportunity to create a valuable derivative data product. Companies such as Urban Engines have demonstrated an ability to add value to the data derived from transit fare transactions. The challenge for the public sector is to, for the societal good, leverage private sector interest by giving access to useful fare transaction data in a manner that protects customer privacy. This challenge is particularly acute in California, where privacy laws make sharing data in a manner that supports the public interest difficult. This paper presents the Metropolitan Transportation Commission's (MTC's) proposed solution to the problem. MTC operates the Clipper® transit fare payment system for the San Francisco Bay Area. In an effort to share usable data that protects customer privacy, MTC developed an anonymizing scheme that is the subject of the present paper. We seek feedback on our approach from the *Data for Good Exchange* community, asking: in seeking a balance between customer privacy and usability, does the scheme go too far in either direction? And, should we take a different anonymizing approach?

## 1. INTRODUCTION

The Metropolitan Transportation Commission (MTC) is the transportation planning, financing, and coordinating agency for the nine-county San Francisco Bay Area – the San Francisco Bay shoreline comprises San Francisco, San Mateo, Santa Clara, Alameda, Contra Costa, Solano, Napa, Sonoma, and Marin Counties. MTC began an effort to introduce a single fare payment system across the many transit operators that serve the region in 1998. Today the Clipper® card can be used on twenty Bay Area transit providers. The service handles around 800,000 transactions each week day and settles over $40 million dollars each month.

As noted above, in addition to operating Clipper®, MTC serves as the region's transportation planning agency. As the Clipper® card gained wider adoption, the transportation planners at MTC sought detailed transaction data to better understand the travel behavior of Bay Area residents. Around the same time, private sector interest in the data increased, in particular from the firm Urban Engines, who sought the data to support their software business. To MTC's planners, the highest value aspect of the Clipper® transaction data is individual trajectories through the transportation network. This data has the potential to reveal interesting and important behaviors, including preference for rail (rather than bus), disdain for transferring, seasonal variation, and day-to-day route variation.

Importantly, MTC's planning staff did not want to be the sole customers of the prospective Clipper® data product. Rather, we sought to create a product that was useful to us that we could share with others.

Though housed in the same agency, California's privacy laws – California Streets and Highways Code Section 31490 in particular – make sharing the data complicated. MTC's planning, Clipper®, and legal personnel set about on an effort to create a useful data product that protected customer privacy and could be broadly distributed. The remainder of the paper describes our work.

## 2. PROBLEM

MTC sought to share useful Clipper® data in a manner that protects customer privacy. In this case, the primary customers of the data, MTC's planners, defined "useful" as containing individual trajectories through the transportation network across a full day, with each day in each year represented in the data. California's privacy laws governing electronic payment systems require that any data product that can be shared with MTC's planners can be shared more widely (i.e., MTC's planners receive no preferential access), in particular with private sector actors interested in leveraging the data for business ventures.

## 3. SOLUTION

MTC's planning, Clipper®, and legal teams iteratively experimented with solutions that attempted to balance protecting customer privacy and retaining the data's usefulness. We were guided by our collective judgment and settled on the following anonymizing scheme:

- All personally identifiable information is held in a separate database that was not joined, examined, or considered as part of our effort and no personally identifiable information is present in the anonymous data.

- Each Clipper® card's unique serial number is replaced with a pseudo-random identification field that persists for a single "circadian" day – the data is separated from 3 am to 3 am, as this represents a more logical interval of transit movements than from midnight to midnight. The persistence of the random ID through the day allows for the analysis of a full day's movement through the transportation network.

- A sample of 50 percent of unique cards is selected for each circadian day. This obscures habitual behavior that, when deviated from, may suggest (when the habitual movements are very uncommon) individual travel patterns – confirming patterns and/or the identity of specific individuals would require other (e.g., observation, a personal relationship, etc.) information. For example, if a single card makes a unique movement each day, the absence of that movement could reveal characteristics of a customer the customer may not want publically known.
- Of the four or five Mondays in a month of data, three are randomly selected. The same approach holds for each day of the week. This further obscures habitual patterns. When the card sampling and day-of-week sampling is combined, each card has a $(0.50 * 3/5 = )$ 30.0 to $(0.50 * 3/4 = )$ 37.5 percent chance of being included in each daily data set, which means the chance of a single card being selected for a full work week is less than $(0.375^5 = 0.7\%)$ one percent. The day-of-week field is retained, which facilitates day-of-week-specific analyses (e.g., are Mondays different than Wednesdays?).
- Each date is replaced with a unique, random number. The combination of year (which is retained), month (also retained), day-of-week, and unique date number allows for the identification of transactions which occurred on a single day, while obscuring the identity of the exact day. This suppression reduces the possibility of pairing surveillance video with the data to identify individual travelers.
- Each time stamp is truncated to the nearest 10 minutes. This further reduces the likelihood of pairing the data with surveillance video to identify an individual. To maintain the correct sequence of transactions that occur within identically-labeled time intervals, a trip sequence identifier is added to the data. This identifier simply counts trips in temporal order made during the circadian day.

Table 1 on the following page presents the full data schema.

Each individual requesting the data must sign a non-disclosure agreement prior to working with the data; the form can be found here: https://mtcdrive.box.com/anonymous-clipper. Among other provisions, the agreement prohibits users from attempting to identify individuals with the data.

### 3.1. INITIAL CRITICISMS OF THE FORMAT

MTC staff solicited criticisms of the data product from internal and external users. These observations included the following:

- Replacing card unique serial numbers with an identifier that only persists for a 24-hour period limits use of the data to identify trends over multiple days in a week, such as the use of a card by weekly commuters traveling to local airports, or on parallel systems.
- Obfuscation of the date was critiqued in two ways. First, some users were explicitly interested in seeing variations in ridership on days that saw special events, such as station closures or large sporting events. Second, users suggested that it might be straightforward to reverse the exact date obfuscation using outside data sources.
- The limitations of route information surprised some users who were expecting to see route information for each transaction (only select transit providers configure the payment system to record route).
- The monthly data sets, which contain about seven million transaction records, are too large for many users.

### 3.2. SUCCESS TO DATE

To date, 18 months of data, each containing the aforementioned seven million transaction records, have been distributed to about 20 users. The scheme and resulting usability of the data was the subject of a paper presented at the 2016 Annual Meeting of the Transportation Research Board (Erhardt, 2016).

### 4. CONCLUSION

We have brought this paper to *Data for Good Exchange* to both share our work and get feedback from the community. Specifically, we seek to learn the following:

- Are attendees aware of other efforts to create open, transaction-scale, anonymized transit fare payment data?
- Do you think our data is useful?
- Did our anonymizing scheme go too far in obscuring the data in the name of protecting customer privacy? Or not far enough?
- Should we exchange different sampling and obfuscation tools in the anonymization process? For example, would you prefer to see a smaller sample size and more accurate time information? Or the other way around?
- Should we approach anonymizing in a different manner?
- Are you aware of anonymizing standards in this or other spaces that we could use to guide our efforts?
- Do you see value in developing an anonymizing standard in this space?

### 4.1. MOVING FORWARD

We see a business need for a smarter, automated connection between all manner of databases that contain sensitive information and end users seeking anonymized and/or aggregations of said data that do not contain sensitive information. We think this can be accomplished by first exposing the full database schema and sample data to end users and allowing users to create queries – from the full dataset – extracting the information they seek. Intelligent interfaces can then assess the queries and resulting data for the likely presence of sensitive information, with human intervention triggered when the likelihood exceeds some threshold. This human intervention could be used to further train the intelligent interface. Such an approach would remove the types of "usability" assessments made by us in the present paper and allow for more efficient delivery of useful data to a broader array of stakeholders.

### 5. ACKNOWLEDGEMENTS

**Table 1: Schema for Anonymized Data**

| Field Name | Data Type | Example | Description |
| --- | --- | --- | --- |
| ClipperCardID | varbinary | D88268EA105… | Anonymized Clipper® card identifier |
| TripSequenceID | int | 2 | Circadian Day Trip Sequence |
| AgencyID | int | 1 | Transit Agency Integer |
| AgencyName | char | AC Transit | Transit Agency Name |
| RouteID | int | 300 | Route Integer |
| RouteName | char | F | Route Name |
| FareAmount | money | 0 | Fare |
| PaymentProductID | int | 119 | Payment Product Integer |
| PaymentProductName | char | AC Transit Adult local pass | Payment Product Name |
| TagOnTime_Time | time | 17:30:00 | Boarding Tag Time |
| TagOnLocationId | int | 2 | Boarding Tag Location Integer |
| TagOnLocationName | char | Transbay Terminal | Boarding Tag Location Name |
| TagOffTime_Time | time | 20:20:00 | Alighting Tag Time |
| TagOffLocationId | int | 15 | Alighting Tag Location Integer |
| TagOffLocationName | char | Millbrae (Caltrain) | Alighting Tag Location Name |
| Year | int | 2013 | Transaction Year |
| Month | int | 10 | Transaction Month (1 is January) |
| DayOfWeekID | int | 4 | Transaction Day of Week Integer |
| DayOfWeek | char | Wednesday | Transaction Day of Week Name |
| RandomWeekID | int | 6 | Random Integer that Identifies a Unique Day |